\documentclass{article}
\usepackage{arxiv}
\usepackage{cite}
\usepackage{amsmath,amssymb,amsfonts}
\usepackage{algorithmic}
\usepackage{graphicx}
\usepackage{textcomp}
\usepackage{tabularx}
\usepackage{hyperref}       
\def\BibTeX{{\rm B\kern-.05em{\sc i\kern-.025em b}\kern-.08em
    T\kern-.1667em\lower.7ex\hbox{E}\kern-.125emX}}

\hypersetup{
	pdftitle={Leveraging Clinical Text and Class Conditioning for 3D Prostate MRI Generation},
	pdfsubject={eess.IV},
	pdfauthor={Emerson P.~Grabke, Babak Taati, Masoom A.~Haider},
	pdfkeywords={Generative AI, Large Language Model Adapters, Medical Image Generation, Magnetic Resonance Imaging, Prostate Cancer},
}
\renewcommand{\shorttitle}{Leveraging Clinical Text and Class Conditioning for 3D Prostate MRI Generation}
\renewcommand{\undertitle}{© 2025 IEEE.  Personal use of this material is permitted.  Permission from IEEE must be obtained for all other uses, in any current or future media, including reprinting/republishing this material for advertising or promotional purposes, creating new collective works, for resale or redistribution to servers or lists, or reuse of any copyrighted component of this work in other works.}
\title{Leveraging Clinical Text and Class Conditioning for 3D Prostate MRI Generation}
\usepackage{authblk}

\author[1,2,3]{%
	Emerson P.~Grabke\thanks{\texttt{e.grabke@mail.utoronto.ca}\qquad Corresponding Author}%
}
\author[1,3,4,5]{%
	Babak Taati\thanks{\texttt{taati@cs.utoronto.ca}\qquad \qquad}%
}
\author[1,2,6]{%
	Masoom A.~Haider\thanks{\texttt{m.haider@utoronto.ca}\qquad \qquad Senior Author}%
}
\affil[1]{Institute of Biomedical Engineering, University of Toronto}
\affil[2]{Lunenfeld-Tanenbaum Research Institute, Mount Sinai Hospital}
\affil[3]{KITE Research Institute, Toronto Rehabilitation Institute, University Health Network}
\affil[4]{Department of Computer Science, University of Toronto}
\affil[5]{Faculty Affiliate of the Vector Institute, Toronto}
\affil[6]{Joint Department of Medical Imaging, University of Toronto, Princess Margaret Hospital, and Sinai Health systems}

\begin{document}
\maketitle
\begin{abstract} 
	Objective: Latent diffusion models (LDM) could alleviate data scarcity challenges affecting machine learning development for medical imaging. However, medical LDM strategies typically rely on short-prompt text encoders, nonmedical LDMs, or large data volumes. These strategies can limit performance and scientific accessibility. We propose a novel LDM conditioning approach to address these limitations. Methods: We propose \textbf{C}lass-\textbf{C}onditioned \textbf{E}fficient \textbf{L}arge \textbf{L}anguage model \textbf{A}dapter (CCELLA), a novel dual-head conditioning approach that simultaneously conditions the LDM U-Net with free-text clinical reports and radiology classification. We also propose a data-efficient LDM pipeline centered around CCELLA and a proposed joint loss function. We first evaluate our method on 3D prostate MRI against state-of-the-art. We then augment a downstream classifier model training dataset with synthetic images from our method. Results: Our method achieves a 3D FID score of 0.025 on a size-limited 3D prostate MRI dataset, significantly outperforming a recent foundation model with FID 0.070. When training a classifier for prostate cancer prediction, adding synthetic images generated by our method during training improves classifier accuracy from 69\% to 74\% and outperforms classifiers trained on images generated by prior state-of-the-art. Classifier training solely on our method's synthetic images achieved comparable performance to real image training. Conclusion: We show that our method improved both synthetic image quality and downstream classifier performance using limited data and minimal human annotation. Significance: The proposed CCELLA-centric pipeline enables radiology report and class-conditioned LDM training for high-quality medical image synthesis given limited data volume and human data annotation, improving LDM performance and scientific accessibility.
\end{abstract}


\section{Introduction}
\label{sec:introduction}
Machine learning (ML) for medical image interpretation is a rapidly expanding field intent on improving imaging-based healthcare. However, recent ML advancements outside of medical imaging require training data in higher orders of magnitude than their predecessors. These advances cannot translate easily to the medical imaging space due to labeled data scarcity challenges including patient privacy and the cost of expert annotation. Recent medical imaging literature has seen an increase in latent diffusion model (LDM) studies~\cite{kazerouniDiffusionModelsMedical2023}. Artificially increasing ML training volume with synthetic, generated data provides a possible solution to labeled training data scarcity~\cite{guoMAISIMedicalAI2025}.

Developing LDMs for medical imaging poses a few challenges. Training LDMs from scratch require large data volumes. Stable Diffusion~\cite{rombachHighResolutionImageSynthesis2022} trained on 400~million nonmedical image-text pairs~\cite{schuhmannLAION400MOpenDataset2021} and subsequently on 5.85~billion pairs~\cite{schuhmannLAION5BOpenLargescale2022}. These data volumes are inaccessible to most healthcare institutions due to data scarcity challenges. Nonmedical 2D LDMs have been successfully fine-tuned on 2D medical images~\cite{bluethgenVisionLanguageFoundation2024}, however this strategy cannot easily extend to 3D. Labeled data conditioning can increase data efficiency during LDM training~\cite{yellapragadaPathLDMTextConditioned2024}. Similar to Stable Diffusion, some medical LDMs have used the pre-trained CLIP text encoder for LDM conditioning~\cite{rombachHighResolutionImageSynthesis2022}. However, research has shown that the CLIP text encoder mostly emphasizes the first 20 tokens~\cite{zhangLongCLIPUnlockingLongText2025}. Consequently, these medical LDM studies typically use only short, curated medical phrases or keywords with either the existing~\cite{choMediSynTextGuidedDiffusion2024} or fine-tuned~\cite{bluethgenVisionLanguageFoundation2024} CLIP text encoder rather than using detail-rich clinical reports. These strategies require extra data annotation, preprocessing, or volume that can be challenging for a single institution. This consequently limits the scientific accessibility of medical LDM development.

We propose \textbf{C}lass-\textbf{C}onditioned \textbf{E}fficient \textbf{L}arge \textbf{L}anguage Model \textbf{A}dapter (CCELLA), a text adapter extending ELLA~\cite{huELLAEquipDiffusion2024}, to improve LDM performance for synthetic medical image generation in a challenging medical imaging domain. CCELLA processes the text embedding generated by a large language model (LLM) from clinical reports prior to dual-head conditioning of the LDM U-Net. CCELLA performs two simultaneous functions: one head aligns the text embedding with the image latent space; the second head provides radiologist-derived class information. These functions have independently improved LDM performance outside of medical imaging~\cite{huELLAEquipDiffusion2024,nicholImprovedDenoisingDiffusion2021}. This joint function is hypothesized to provide two advantages: 1) to assert that the LDM receives conditioning about the presence or absence of suspicious lesions in the image, 2) to inform CCELLA which text features should be emphasized for LDM conditioning. Performing both functions in one adapter is hypothesized to improve LDM performance beyond performing both functions independently. We also propose a modified loss function that simultaneously trains the LDM to produce high quality medical images and to appropriately extract radiology classification information from the clinical report text.

Finally, we propose an LDM training pipeline around CCELLA that uses limited human data annotation and a training data volume reasonably accessible to a single healthcare institution. The pipeline intent is to improve the scientific accessibility of LDMs for medical imaging despite limited available data volumes. The pipeline functions by exploiting the clinical report and labeling already performed by radiologists during clinical routine and by reducing training requirements with publicly available, pre-trained components~\cite{guoMAISIMedicalAI2025,chungScalingInstructionFinetunedLanguage2024}.

We evaluated this pipeline within the prostate MRI domain. Radiologists score prostate MRI using the Prostate Imaging -- Reporting and Data System (PI-RADS) guidelines~\cite{PIRADSV21PIRADS}. This domain was selected for its clinical interpretation difficulty, its moderate inter-reader variability between clinical experts, and the relatively low availability of publicly available domain-specific data~\cite{stabileFactorsInfluencingVariability2020,sahaArtificialIntelligenceRadiologists2024}. We performed ablations to assess how CCELLA improves LDM performance and compared performance differences between variants of CCELLA. To validate utility of the proposed LDM pipeline for dataset augmentation, this study demonstrated that the synthetic data generated by our pipeline improved classifier performance when interpreting prostate MRI for cancer presence.

In summary, this paper proposes the following:
\begin{itemize}
	\item A novel dual-head adapter for simultaneous alignment of medical text features extracted by a nonmedical LLM to the medical imaging space and PI-RADS class extraction from medical text.
	\item A LDM training pipeline for improved radiology class-conditioned synthetic medical image generation that is trainable with a reasonable data volume.
	\item A method for augmenting datasets to improve downstream ML tasks using LDM-generated synthetic images with limited data annotation beyond clinical routine.
\end{itemize}

\section{Related Work}
\label{sec:relatedwork}
\subsection{Latent Diffusion Models}
LDMs are a type of generative model with two stages: a vector-quantized variational autoencoder (VQ-VAE) and a denoising U-Net~\cite{rombachHighResolutionImageSynthesis2022}. The VQ-VAE compresses the image to a lower-dimensional representation in a discretized latent space~\cite{vandenoordNeuralDiscreteRepresentation2017}. This reduces computational complexity increasing training and inference speeds, and aligns the image representation to a normal distribution. The denoising U-Net learns the reverse process to a Markov chain~\cite{hoDenoisingDiffusionProbabilistic2020}. During LDM training, the Markov chain forward process gradually adds Gaussian noise to the latent image for each of many (e.g. 1000) timesteps. The denoising U-Net is trained to reverse the forward process by predicting the noise added between the true latent image and a randomly selected timestep. Denoising Diffusion Probabilistic Models (DDPM)~\cite{hoDenoisingDiffusionProbabilistic2020} refers to a diffusion-based generative approach originally formulated for pixel space, with LDMs applying the DDPM framework within a compressed latent space. On inference, the synthetic latent image begins as random Gaussian noise. The U-Net then denoises the latent from timestep 1000 to 0. The VQ-VAE then decodes the resultant latent image to yield a synthetic image.

\subsection{Conditioning and Medical Diffusion Models}
LDMs have shown promising results for synthetic medical and nonmedical image generation~\cite{kazerouniDiffusionModelsMedical2023}. Medical LDMs typically require large data volumes to train from scratch, or strategic data-efficient methods to adapt pretrained LDMs. LDM conditioning with additional information (e.g. segmentation maps, text, or PI-RADS class) is a critical approach for improved image generation performance~\cite{wildeMedicalDiffusionBudget2024,xuMedSynTextguidedAnatomyaware2024,guoMAISIMedicalAI2025}.

Conditioning enables denoising U-Net behavior control and typically takes up to three forms. The first is timestep embedding, informing the LDM of its stage in the denoising process. Similar to positional encoding in transformers~\cite{vaswaniAttentionAllYou2017}, sinusoidal timestep encoding conditions the LDM through addition with the latent image. The second is prompt embedding, informing the LDM of additional information during denoising. Prompting data (e.g. images, text) are separately encoded and condition the LDM through concatenation with the latent image (e.g. for image prompting) or by multi-layer cross-attention within the denoising U-Net (e.g. for text prompting). The third is class conditioning, which expands upon the timestep embedding by adding or concatenating a vector with class information to the timestep embedding~\cite{nicholImprovedDenoisingDiffusion2021}. Conditioning methods can occur in parallel. For example, MAISI utilized prompting with segmentation masks, and included both voxel spacing and body region for class conditioning~\cite{guoMAISIMedicalAI2025}.

LDM text conditioning typically relies on the Contrastive Language-Image Pre-Training (CLIP) text encoder~\cite{radfordLearningTransferableVisual2021}, trained to encode text to a shared image-text latent space. The CLIP text encoder is limited by a 77 text token input restriction, with literature documenting an effective length of under 20 tokens~\cite{zhangLongCLIPUnlockingLongText2025}. Thus, commercial LDMs typically restrict text prompting to keywords or short phrases. Conversely, clinical reports are typically longer and rich in free-text information (e.g. location, disease severity) that could improve synthetic image realism~\cite{yellapragadaPathLDMTextConditioned2024}. Medical imaging LDM text prompting strategies include crafting short phrases~\cite{saeedBiparametricProstateMR2024}, training new embeddings for a frozen CLIP text encoder~\cite{wildeMedicalDiffusionBudget2024}, fine-tuning the CLIP text encoder on medical data~\cite{bluethgenVisionLanguageFoundation2024}, or training new LDMs with a domain-relevant medical LLM~\cite{bluethgenVisionLanguageFoundation2024,xuMedSynTextguidedAnatomyaware2024}.

The CLIP text token restriction has prompted innovation in adapting large language models (LLMs) for LDM use. Notably, the recent Efficient Large Language model Adapter (ELLA) demonstrated improvement in both LDM image generation and long-prompt following~\cite{huELLAEquipDiffusion2024}. ELLA adapts text encoded by the T5 XL LLM using a chain of six Timestep-aware Semantic Connector (TSC) blocks to generate the final embedding to condition the LDM. TSC block design details can be found in~\cite{huELLAEquipDiffusion2024}. ELLA leverages the power of LLMs for text feature extraction without LLM or LDM retraining. 

PathLDM~\cite{yellapragadaPathLDMTextConditioned2024} is the most similar in motivation to our work, recognizing the value of full-length clinical text, class-based information, and automated ground-truth labeling combined with the LDM-related limitations of the CLIP text encoder. PathLDM developed an automated text prompt generation workflow for synthetic 2D breast cancer histopathology slide generation. This workflow extracted two pathology-related probabilities (``Tumor'' and ``TIL'') with pretrained models, binarized them as ``high'' or ``low'', and summarized each clinical report to under 75 words using the GPT-3.5 LLM~\cite{brownLanguageModelsAre2020}. PathLDM also replaced the CLIP text encoder with a pretrained, histopathology-specific CLIP text encoder variant~\cite{huangLeveragingMedicalTwitter2023}. The final text prompt combined the two class-based keywords and summarized report in the form ``[high/low] Tumor, [high/low] TIL, [report summary]'', thereby conditioning PathLDM with both classifier and key text report data.

Different from PathLDM, our method:
\begin{itemize}
	\item Uses text from the original report instead of a shorter, LLM-generated summary;
	\item Trains the class extractor simultaneously with the LDM, rather than relying on a nonmedical, pretrained solution;
	\item Adapts a pretrained nonmedical LLM for medical use instead of using a medical CLIP text encoder variant;
	\item Performs class conditioning using timestep instead of text prompt, enabling class-specific text adapter training.
\end{itemize}

Compared to ELLA, our method:
\begin{itemize}
	\item Modifies each TSC block by adding PI-RADS class extractor layers;
	\item Focuses additionally on class conditioning rather than solely on adapting the text embedding to the latent space;
	\item Focuses on 3D medical instead of 2D nonmedical images;
	\item Trains the LDM U-Net from scratch alongside the adapter rather than using a pretrained LDM U-Net.
\end{itemize}

\section{Methods}
\label{sec:methods}
\subsection{Dataset}
This retrospective study used axial T2-weighted prostate MRI of men at high-risk for prostate cancer. An institutional research ethics board approved this study and informed consent was waived. The complete dataset used 4153 images from one healthcare institution plus 1658 images from two publicly available datasets (1500 from the PI-CAI Grand Challenge~\cite{sahaArtificialIntelligenceRadiologists2024}, 158 from Prostate158~\cite{adamsProstate158Expertannotated3T2022}) for a total of 5811 images. Healthcare institution inclusion criteria were referral for prostate MRI, being treatment naïve, and study date between April 2009 and August 2021. Exclusion criteria were incomplete or non-diagnostic image quality. Healthcare institution DICOM images were converted to NIFTI with patient information removed from image metadata prior to this study. Radiology reports were preprocessed to remove identifying information (e.g. patient name, doctor initials). Public dataset inclusion and exclusion criteria have been published~\cite{sahaArtificialIntelligenceRadiologists2024,adamsProstate158Expertannotated3T2022}. The public datasets were already anonymized.

Clinical reports were available for 4150 studies from the healthcare institution. Neither public dataset included clinical reports with their images. Images from the healthcare institution that were acquired prior to the latest PI-RADS update were reviewed and assigned a new PI-RADS score per the updated guidelines~\cite{PIRADSV21PIRADS}. All patients had either a PI-RADS classification (positive = PI\nobreakdash-RADS$\ge$3) or histopathology results from within 6 months of the MRI (positive = ISUP$\ge$2). Training set images missing a radiology classification instead used the histopathology result. A hold-out test set of 740 images was randomly selected on a per-case level and used to assess the results in this study. 

\subsection{Latent Diffusion Model Implementation}
The proposed LDM pipeline used two pretrained components. The first is the MAISI LDM VQ-VAE, pretrained on 3D CT and MRI images~\cite{guoMAISIMedicalAI2025}. The second is the FLAN-T5 XXL LLM~\cite{chungScalingInstructionFinetunedLanguage2024}. The base LDM in this study used the frozen MAISI VQ-VAE with a learnable 4-channel, 4-layer denoising U-Net that concatenated image voxel spacings to the timestep embedding. The frozen FLAN-T5 XXL text encoder was used to encode the clinical report text.

\begin{figure*}
	\centering
	\includegraphics[width=\linewidth]{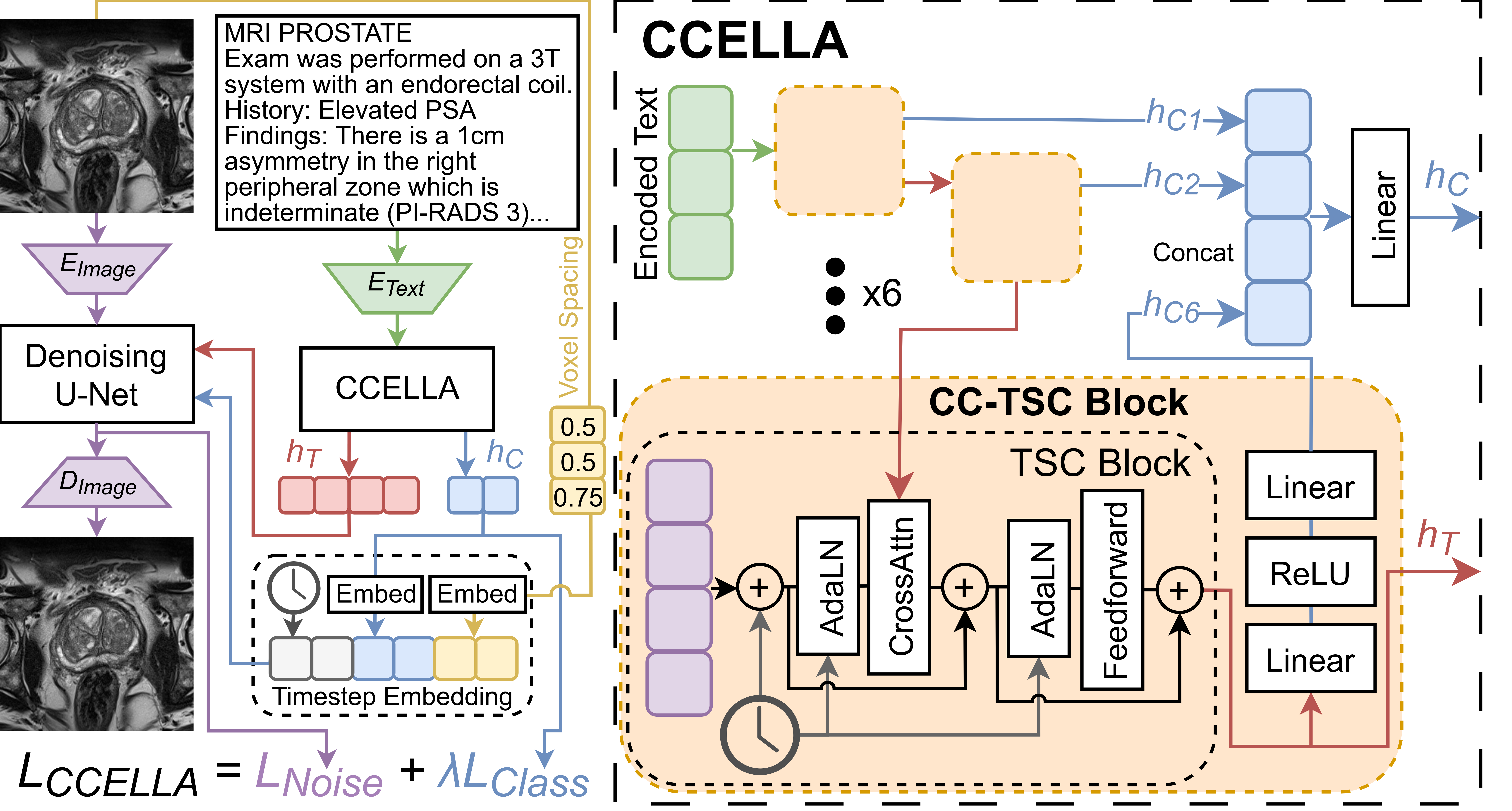}
	\caption{Overview of CCELLA-centric pipeline (left) and CCELLA adapter (right). Image encoder ($E_{Image}$) and decoder ($D_{Image}$) from~\cite{guoMAISIMedicalAI2025}. Text encoder ($E_{Text}$) from~\cite{chungScalingInstructionFinetunedLanguage2024}. CCELLA consists of six CC-TSC blocks with classifier ($h_C$) and aligned text embedding ($h_T$) heads. TSC block originally from~\cite{huELLAEquipDiffusion2024}. Pipeline input includes the radiology report text for training and inference, with image and radiology classification additionally included for training alone. Example text redacted for privacy.}
	\label{fig1}
\end{figure*}

CCELLA extends the ELLA adapter from~\cite{huELLAEquipDiffusion2024} by adapting encoded text for both deep cross-attention and timestep embedding conditioning. CCELLA begins with six class-conditioning TSC (CC-TSC) blocks. These blocks each augment the ELLA adapter TSC block by adding a separate classifier head of two fully-connected layers of hidden size 256 and output size 2 (MLP block) to the end of the block. CCELLA uses the separate classifier head outputs to produce a class prediction from the encoded text. All CC-TSC classifier head outputs are concatenated for input to a final fully-connected layer, which produces a final class prediction of size 2 to be concatenated with the timestep embedding for LDM conditioning. The image voxel spacing was additionally concatenated to the timestep embedding as per the foundation MAISI model~\cite{guoMAISIMedicalAI2025}. The output of the final CC-TSC text embedding head conditioned the denoising U-Net through cross-attention. Figure \ref{fig1} illustrates the CCELLA adapter design as well as the full LDM pipeline dataflow. Class conditioning was intentionally implemented as a separate timestep embedding vector instead of as additional text prompt words as performed in PathLDM~\cite{yellapragadaPathLDMTextConditioned2024}. Our dual text- and classifier-head implementation enabled direct training of the CCELLA adapter using radiologist classification rather than relying on implicit adapter training for improved image generation alone as performed by ELLA~\cite{huELLAEquipDiffusion2024}.
To train CCELLA with the denoising U-Net, a new joint loss function is proposed:
\begin{equation}L_{CCELLA} = L_{NoisePred} + \lambda L_{Class} \label{eq_loss}\end{equation}

The intent of the loss function in \eqref{eq_loss} was to emphasize both LDM image generation through noise prediction and PI-RADS class extraction from the reference text. Specifically, $L_{NoisePred}$ in \eqref{eq_loss} trains U-Net noise prediction by comparing the forward diffusion process noise with the noise predicted by the denoising U-Net, for the purpose of image generation. $L_{Class}$ in \eqref{eq_loss} trains class prediction by comparing the CCELLA class prediction ($h_C$ in Figure \ref{fig1}) with the ground truth radiology class. Because the class predictor head receives information from each CC-TSC block, $L_{Class}$ enables direct training of the CCELLA adapter based on information it should extract from the clinical text. This was hypothesized to improve the radiology text feature extraction quality, thereby improving the downstream image generation performance.

We used L1 loss for $L_{NoisePred}$ and focal loss for $L_{Class}$, with class loss weighting \(\lambda = 10^{-4}\). L1 loss was selected for its frequent use in LDM noise prediction~\cite{rombachHighResolutionImageSynthesis2022,guoMAISIMedicalAI2025}. Focal loss was selected for its success in the PI-CAI challenge~\cite{sahaArtificialIntelligenceRadiologists2024}. Class loss weighting was determined empirically after training on a subsample of the full training set.

\subsection{CCELLA Training and Evaluation}
MAISI, ELLA, and PathLDM have already demonstrated success against state-of-the-art models~\cite{guoMAISIMedicalAI2025,huELLAEquipDiffusion2024,yellapragadaPathLDMTextConditioned2024}. Specifically, MAISI demonstrated improved Fréchet Inception Distance (FID) in all 2D planes compared to a pretrained, medical generative adversarial network~\cite{guoMAISIMedicalAI2025}. ELLA demonstrated qualitative and quantitative improvement when added to an existing commercial LDM~\cite{huELLAEquipDiffusion2024}. PathLDM~\cite{yellapragadaPathLDMTextConditioned2024} demonstrated qualitative and quantitative improvement over Stable Diffusion~\cite{rombachHighResolutionImageSynthesis2022} for generating text-conditioned histopathology images. Thus, we assessed the CCELLA-centric pipeline against: 1) the MAISI LDM (MAISI) without text or class conditioning, 2) MAISI with radiology classification timestep conditioning (RADMAISI), 3) MAISI with text conditioning using ELLA to adapt FLAN-T5 XXL (ELLAMAISI), 4) PathLDM with minor modification to accommodate the different data modality and ethics requirements in this study. To implement PathLDM, three adjustments were required. Firstly, PathLDM was developed for 2D image generation. To accommodate 3D image generation, the model was modified to reuse the MAISI VQ-VAE and to train a 3D denoising U-Net from scratch. Secondly, PathLDM used the online, publicly available GPT 3.5~\cite{brownLanguageModelsAre2020} to summarize clinical text reports into lengths up to 154 tokens. Because patient data processing using online services was not allowed under the ethics agreement for this study, a Python script instead extracted the ``Impression'' section from each text report. This radiologist-summarized report section was then used in lieu of an LLM-generated summary for text conditioning, similar to the work performed by Bluethgen et al.~\cite{bluethgenVisionLanguageFoundation2024}. Thirdly, the histopathology-specific CLIP text encoder used by PathLDM is not relevant for prostate MRI report text, thus the original CLIP text encoder from Radford et al.~\cite{radfordLearningTransferableVisual2021} was used instead. These modifications were not anticipated to negatively affect model performance, rather it may possibly improve performance for this experiment as the radiologist summary may outperform the LLM summary. Encoded text for conditioning the modified PathLDM began with ``High tumor;'' or ``Low tumor;'' depending on its classification, followed by the summarized report.

We then performed an ablation study to assess the individual text and classifier conditioning components of CCELLA as well as the joint loss function. Aforementioned baseline models RADMAISI and ELLAMAISI represent the LDM with classifier or text conditioning alone (respectively). Additional ablations included: 1) an LDM with ELLA text conditioning and separate timestep-based class conditioning using ground truth radiology classification (DECOUPLED), 2) CCELLA using only L1 loss for noise prediction instead of the proposed joint loss in \eqref{eq_loss} (L1ONLY).

Finally, a variant study was performed, comparing the CCELLA design to variants by: 1) placing only one CC-TSC block after a chain of five TSC blocks (LAST), 2) placing one CC-TSC block before five TSC blocks (FIRST), 3) chaining six CC-TSC blocks where class adapter weights were shared across all blocks (SHARED).

All models received image voxel spacing as timestep conditioning as per the foundation MAISI model~\cite{guoMAISIMedicalAI2025}. All models used the DDPM architecture with 1000 time steps. The DDPM architecture was selected for consistency with the foundation MAISI model, enabling better comparison between baseline, ablation, and variant models. Studies without text reports were not excluded, representing a more realistic data scarcity scenario with heterogeneous data. Our pipeline allowed for training on data regardless of associated text report presence. This was achieved by disabling CCELLA classifier head backpropagation for studies without a text report (i.e. setting $\lambda = 0$ in \eqref{eq_loss}) during training. This allowed the denoising U-Net to train on a larger data volume, while also improving the text adapter for studies with text reports.

\textbf{Training details:} All images were center-cropped to a $128mm \times 128mm$ axial plane field of view, normalized and then resampled to $ 256 \times 256 \times 128$ voxel dimensions using Lanczos interpolation~\cite{duchonLanczosFilteringOne1979} prior to LDM training and evaluation. All LDMs were trained for 400 epochs with initial learning rate of \(10^{-4}\), the AdamW optimizer, a second-order polynomial learning rate scheduler, and batch size of 16 on four Tesla V100 GPUs. When training LDMs with text conditioning, images without a text report used an empty string as a proxy report prior to LLM encoding with appropriate attention map. When training CCELLA, \(\lambda\) in \eqref{eq_loss} was set to 0 for these cases so that the classifier portion of the loss function did not update the model for these images. When training PathLDM on cases without text, the encoded text consisted solely of the classifier text component (e.g. ``High tumor;'').

\textbf{Evaluation:} LDM performances were evaluated using FID~\cite{heuselGANsTrainedTwo2017}. FID was calculated in two ways: 1) in 3D using the Med3D ResNet50 pretrained on 23 datasets~\cite{chenMed3DTransferLearning2019}, 2) in 2D for each of the sagittal (sag.), coronal (cor.) and axial (ax.) planes using a ResNet50 pretrained on RadImageNet~\cite{meiRadImageNetOpenRadiologic2022}.

\subsection{Downstream Classification Task}
The trained CCELLA, ELLAMAISI, and PathLDM models each generated one synthetic image for every real training sample using the sample's clinical text. Ground truth for each ELLAMAISI- and PathLDM-generated image was set to the radiologist label from the source training sample (``clinical label''). For CCELLA-generated images, both the clinical label and the CCELLA-generated label (``synthetic label'') were stored for each synthetic image during generation. EfficientNet-b0~\cite{tanEfficientNetRethinkingModel2019} models were then trained in two scenarios for each LDM: 1) synthetic data with clinical labels, 2) real and synthetic data with clinical labels for synthetic images. Two additional scenarios were trained for CCELLA-generated images: 3) synthetic data with synthetic labels, 4) real and synthetic data with synthetic labels for synthetic images. Finally, a baseline classifier was trained on real data only. For each LDM-specific scenario, the same real and synthetic data were used in training (i.e. synthetic data from one LDM was not regenerated between scenarios for that LDM).

\textbf{Training details:} All synthetic images were resampled to the original voxel size of their corresponding real image using Lanczos interpolation~\cite{duchonLanczosFilteringOne1979}. All images were linearly rescaled from 0\%-98\% intensity with clipping then padded to a $128mm \times 128mm \times 96mm$ field of view. Images were augmented during training with random flips, scaling, translations, and rotations. All classifiers were trained on four Tesla V100 GPUs for 16k steps with a batch size of 16, an initial learning rate of \(10^{-4}\), the AdamW optimizer, and a second-order polynomial learning rate scheduler.

\textbf{Evaluation:} Trained EfficientNet-b0 performances were assessed using the area under the receiver operating characteristic curve (AUC), average precision (AP), and classification accuracy on the same unseen test set as the LDM evaluation. Secondary assessment was performed using sensitivity, specificity, positive predictive value (PPV), and negative predictive value (NPV).

\section{Results}
\label{sec:results}
\subsection{Performance Comparison}

\begin{table}
	\scriptsize
	\caption{CCELLA Performance vs Baselines}
	\label{table1}
	\setlength{\tabcolsep}{3pt}
	\begin{tabularx}{\linewidth}{|X|X|X|X|X|}
		\hline
		Model&
		3D FID~$\downarrow$ &
		2D Sag. FID~$\downarrow$ &
		2D Cor. FID~$\downarrow$ &
		2D Ax. FID~$\downarrow$ \\
		\hline
		MAISI&
		0.070&10.03&8.75&13.24\\
		RADMAISI&
		0.077&10.96&9.53&14.10\\
		ELLAMAISI&
		0.080&11.64&10.40&15.20\\
		PathLDM&
		0.046&9.07&7.82&12.02\\
		DECOUPLED&
		0.048&8.33&7.58&11.57\\
		L1ONLY&
		0.051&7.93&7.21&10.94\\
		CCELLA (ours)&
		\textbf{0.025}&\textbf{5.27}&\textbf{4.73}&\textbf{8.09}\\
		\hline
	\end{tabularx}
	\label{tab1}
\end{table}

\begin{figure}
	\centering
	\includegraphics[width=\linewidth]{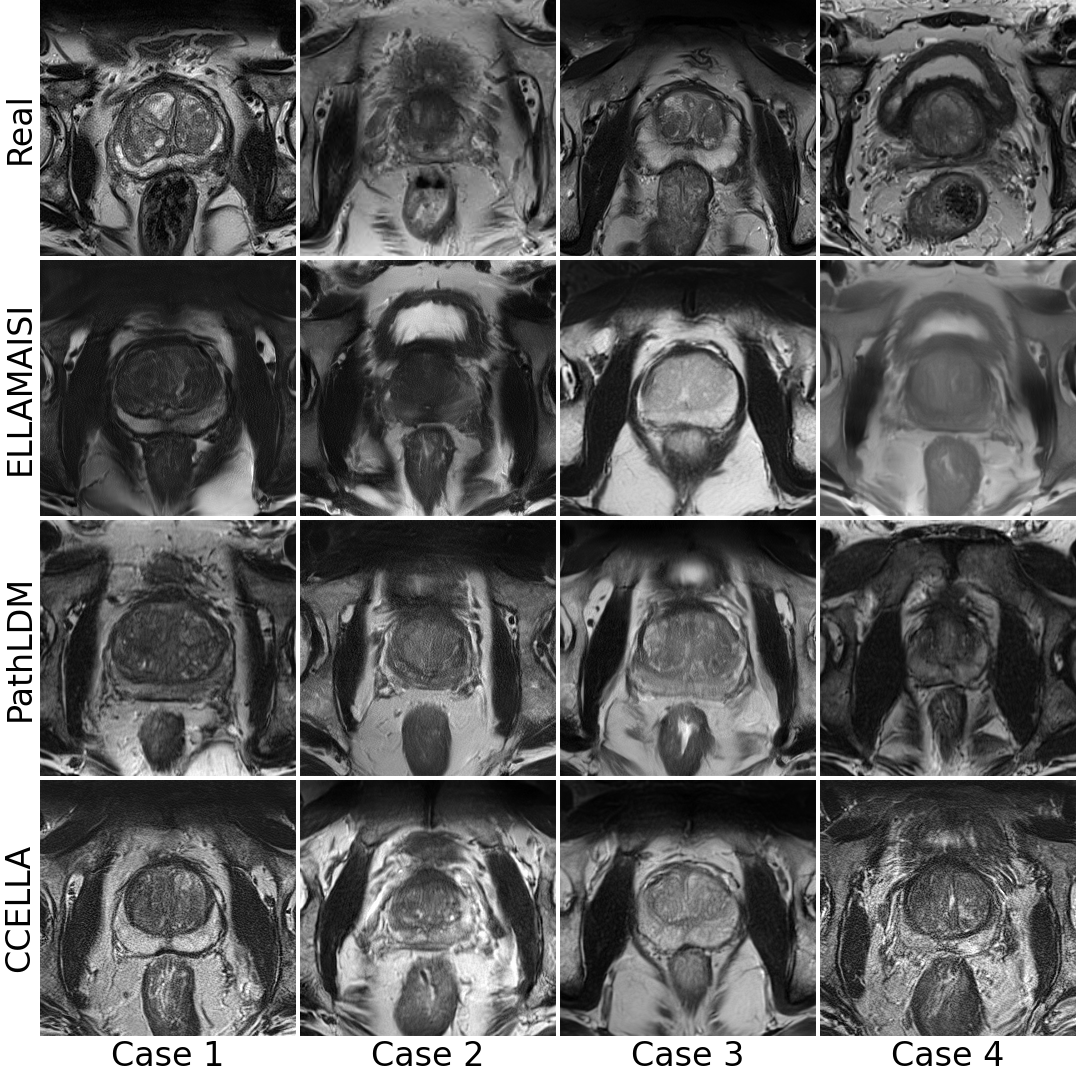}
	\caption{Four real axial T2 prostate MRI images and the synthetic MRI images generated by ELLAMAISI, PathLDM, and CCELLA conditioned on the same corresponding report text.}
	\label{fig2}
\end{figure}

Table \ref{table1} shows the FID performance of CCELLA against its baseline and ablation models. CCELLA outperformed all other models in both 3D and all-axis 2D FID metrics. Providing no conditioning, or only one of timestep or cross-attention conditioning performed worse than providing both. Example images generated by CCELLA, ELLAMAISI and PathLDM can be seen in Figure \ref{fig2}.

\subsection{Ablation Study}
\begin{table}[t]
	\scriptsize
	\caption{Ablation Study: CCELLA Variants}
	\label{table2}
	\setlength{\tabcolsep}{3pt}
	\begin{tabularx}{\linewidth}{|l|X|X|X|X|X|X|X|X|X|}
		\hline
		Model&
		3D FID~$\downarrow$ &
		2D Sag. FID~$\downarrow$ &
		2D Cor. FID~$\downarrow$ &
		2D Ax. FID~$\downarrow$ &
		Acc.~$\uparrow$ &
		Sens.~$\uparrow$ &
		Spec.~$\uparrow$ &
		PPV~$\uparrow$ &
		NPV~$\uparrow$ \\
		\hline
		SHARED&
		0.064&7.84&7.05&11.01&
		\textbf{0.83}&\textbf{0.82}&0.85&0.87&\textbf{0.79}\\
		FIRST&
		0.053&10.60&9.33&14.16&
		0.82&0.79&\textbf{0.86}&\textbf{0.88}&0.76\\
		LAST&
		0.051&8.30&7.16&11.34&
		0.82&0.79&0.85&0.87&0.77\\
		CCELLA&
		\textbf{0.025}&\textbf{5.27}&\textbf{4.73}&\textbf{8.09}&
		0.81&0.77&0.86&0.87&0.75\\
		\hline
		
	\end{tabularx}
	\label{tab2}
\end{table}

Table \ref{table2} shows the FID and confusion matrix performances of CCELLA against its variants. CCELLA outperformed its variants in all FID metrics. Interestingly, CCELLA performed worse than its variants at class prediction from text, with the SHARED variant achieving best class prediction accuracy.

\subsection{Downstream Classification Task}

\begin{table}[t]
	\scriptsize
	\caption{Classification Results when Trained with Combinations of Real Data (R), Synthetic Data with Clinical Labels ($S_R$), and Synthetic Data with Synthetic Labels ($S_S$) Generated by ELLA, PathLDM, or CCELLA}
	\label{table3}
	\setlength{\tabcolsep}{3pt}
	\begin{tabularx}{\linewidth}{|l|l|X|X|X|X|X|X|X|l|}
		\hline
		Scenario    & Model   & AUC           & AP            & Acc.      & Sens.   & Spec.   & PPV           & NPV           & \% Pos \\ \hline
		\(R\)       & -       & 0.76          & 0.73          & 0.69          & 0.67          & 0.72          & 0.75          & 0.63          & 48\%   \\
		\(S_{R}\)       & ELLAMAISI    & 0.77          & 0.77          & 0.67          & 0.60          & 0.77          & 0.77          & 0.60          & 48\%   \\
		\(S_{R}\)       & PathLDM & 0.76          & 0.77          & 0.65          & 0.57          & 0.76          & 0.75          & 0.58          & 48\%   \\
		\(S_{R}\)   & CCELLA  & 0.77          & 0.77          & 0.64          & 0.51          & 0.81          & 0.77          & 0.57          & 48\%   \\
		\(S_{S}\)   & CCELLA  & \textbf{0.80} & \textbf{0.80} & 0.66          & 0.50          & \textbf{0.87} & \textbf{0.83} & 0.58          & 39\%   \\
		\(R+S_{R}\)     & ELLAMAISI    & 0.76          & 0.72          & 0.69          & 0.63          & 0.75          & 0.76          & 0.62          & 48\%   \\
		\(R+S_{R}\)     & PathLDM & 0.76          & 0.78          & 0.70          & 0.64          & 0.76          & 0.77          & 0.63          & 48\%   \\
		\(R+S_{R}\) & CCELLA  & \textbf{0.80} & 0.79          & \textbf{0.74} & \textbf{0.71} & 0.77          & 0.80          & \textbf{0.68} & 48\%   \\
		\(R+S_{S}\) & CCELLA  & 0.79          & 0.79          & 0.71          & \textbf{0.71} & 0.72          & 0.76          & 0.66          & 43\%   \\ \hline
	\end{tabularx}
	\label{tab3}
\end{table}

Table \ref{table3} shows the performance of EfficientNet-b0 classifiers trained for identifying suspicious lesions in prostate MRI trained on any of real (\(R\)), synthetic with clinical label (\(S_{R}\)), and synthetic with synthetic label (\(S_{S}\)) datasets, with synthetic data generated by any of ELLAMAISI, PathLDM, or CCELLA. AUC and AP report diagnostic performance agnostic of prediction threshold. Accuracy (Acc.), sensitivity (Sens.), specificity (Spec.), PPV and NPV were calculated at a decision threshold of 0.5. Training on synthetic or combined synthetic and real data consistently improved AUC and AP in all CCELLA scenarios. This trend did not hold for all ELLA and PathLDM scenarios. All CCELLA and PathLDM metrics were improved or maintained when augmenting real with synthetic data compared to synthetic data alone. Augmenting real with ELLAMAISI-generated synthetic data lowered AUC and AP compared to training on ELLAMAISI synthetic data alone. Compared to training on real data, combining real with CCELLA-generated synthetic images was the only method that improved or maintained all classifier metrics, regardless of synthetic ground truth type. Best classifier metrics in all scenarios were achieved only by CCELLA-derived models, specifically the CCELLA $S_S$ and $R+S_R$ models.

Positive training case proportions were consistent for all models except CCELLA-generated synthetic label models. CCELLA learned to generate a negative label for images without text, resulting in a lower proportion of positive labels (\% Pos) for training for \(S_{S}\). All 468 training dataset cases where CCELLA extracted negative classification from a positive source image occurred when clinical report text was missing. The test set had a positive case percentage of 56\%.

\section{Discussion}
This study sought to introduce a new pipeline for training higher-quality synthetic prostate MRI using a limited and fixed volume dataset with limited supervised labeling. We achieved this by proposing CCELLA and a joint loss function, as well as pretrained component reuse where possible. Our strategy achieved higher performance than baseline LDMs conditioned on nothing, radiology label only, and text only using ELLA. The ablation study sought to understand the mechanism by which CCELLA achieved better performance. In addition to the text-only scenario represented by ELLAMAISI and the class-only scenario represented by RADMAISI, the ablation study trained: 1) DECOUPLED, an LDM receiving independent text conditioning by cross-attention using ELLA and timestep conditioning using the ground truth radiology label, 2) L1ONLY, an LDM fitted with CCELLA but trained without the proposed joint loss. Both RADMAISI and ELLAMAISI were outperformed by DECOUPLED and L1ONLY, suggesting that the combination of text and classifier conditioning (i.e. DECOUPLED) and the combination of cross-attention and timestep conditioning (i.e. L1ONLY) outperforms the use of either individually. DECOUPLED outperformed L1ONLY, indicating that timestep-based class conditioning improves LDM performance over unstructured timestep conditioning. Finally, CCELLA outperformed both DECOUPLED and L1ONLY, confirming the primary study hypothesis that it is not simply the addition of a class label in the timestep embedding (DECOUPLED) nor the extension of the text adapter for timestep conditioning (L1ONLY) that results in CCELLA's success. In combination, these ablation results indicate that the joint training of the classifier with the text adapter encourages CCELLA to extract more clinically relevant information from text encoded by a non-clinically-trained LLM. These ablation results also suggest that pretrained, nonmedical LLMs could be better adapted for medical text processing when trained jointly with clinical labels. Further investigation should be performed to confirm these findings in different medical domains.

CCELLA was also compared against alternate designs in a variant study. Interestingly, these results indicate a near-inverse relationship between diffusion output quality and classifier accuracy, with CCELLA achieving the best in output quality but the worst in classifier accuracy and the SHARED variant achieving the worst in image generation but the best in classifier accuracy. While an initial intuition of these results might suggest a trade-off between classifier and text adapter performances, future investigations should further explore this relationship to understand how it might be better leveraged for both improved LDM and classifier performance.

Figure \ref{fig2} shows four real images alongside CCELLA-, ELLAMAISI-, and PathLDM-generated images using the same corresponding text prompts. Qualitatively, the CCELLA images exhibit natural variances in prostate size and shape while consistently depicting the different anatomical regions within the prostate. CCELLA did not struggle with image brightness and contrast to the same degree as ELLAMAISI and PathLDM for the same prompts. ELLAMAISI exhibited occasional difficulty in defining prostate regions, showing clear, artificial smoothness in Figure~\ref{fig2} Case 4. While PathLDM generated natural prostate MRI features more consistently than ELLAMAISI, it still exhibited some difficulties with artificial smoothness and contrast in Cases 2 and 4 (respectively) compared to CCELLA. Ultimately, these observed features follow the 3D FID pattern in Table~\ref{table1}, where PathLDM outperformed ELLAMAISI, and CCELLA outperformed both PathLDM and ELLAMAISI. The qualitative improvements of CCELLA over ELLAMAISI reaffirm the CCELLA adapter design choice to use class conditioning as a directly trainable output.

Finally, this study trained downstream classifier models for prostate cancer prediction from MRI using real data and synthetic data generated by any of ELLAMAISI, PathLDM, and CCELLA. Consistent with the LDM performances in Table~\ref{table1}, classifiers trained with CCELLA-generated data outperformed those trained with PathLDM data, which in-turn outperformed models trained with ELLAMAISI data. Only the CCELLA \(R+S_R\) and \(R+S_S\) classifiers performed equally or better than the \(R\) classifier on all metrics, with CCELLA \(R+S_R\) outperforming \(R\) on all confusion matrix metrics. These results empirically validate the quantitative results from Table~\ref{table1} that CCELLA improves generated image quality beyond its baseline comparison LDMs. These results also validate the hypothesis that synthetic data can successfully augment a classifier model training dataset, with the caveat that not all synthetic data improves classifier performance as evidenced by failure of PathLDM-generated and ELLAMAISI-generated images to consistently maintain or improve downstream classifier metrics. These results also align with recent literature that generate ground truths alongside data samples. The CCELLA-trained $S_S$ classifier also performed comparably to \(R\) with higher AUC and AP but lower accuracy, suggesting that the CCELLA-generated synthetic data may be as useful as real data when training downstream machine learning tasks after classifier threshold tuning. These results are promising, however the synthetic data was generated from the same distribution as the real data. Future work should seek to investigate these results further by generating synthetic data for training with real data of different distribution.

When investigating CCELLA-specific classifiers a few trends were noted. Interestingly and in contrast to expectation, the \(S_S\) classifier outperformed not only the \(S_R\) classifier in all metrics except sensitivity, but also the \(R\) classifier in multiple metrics (AUC, AP, specificity, PPV). Seemingly paradoxically, augmenting the real dataset with synthetic data in \(R+S_R\) resulted in better performance than augmenting in \(R+S_S\) in all metrics except AP and sensitivity where they were equal. This likely occurred for three reasons. Firstly, only 3701 of the 5071 training samples had clinical report text. Because null text reports generated negative class predictions, adding synthetic data with synthetic labeling in \(R+S_S\) resulted in a lower positive case percentage than using real labels in \(R+S_R\) (43\% versus 48\%). It therefore makes retrospective sense that \(R+S_S\) would exhibit more difficulty than \(R+S_R\) in adapting to a test set with higher positive case percentage (56\%), resulting in lower accuracy (0.71 versus 0.74). Secondly and related to the first, the training cases without radiology text belonged to both positive and negative classes. Even though the joint classifier loss was removed for these cases, the LDM would still have learned to generate both positive and negative images when presented with a null prompt. This in-turn could have affected the performance of classifiers trained with synthetic data. Thirdly, the CCELLA LDM did not receive the real image as input during inference. However, the ground-truth radiology classification was based on the radiologist re-interpretation of the full image under the most recent PI-RADS standard instead of simply extracting the classification in the clinical text. Consequently, the ground truth re-classification may have lost its meaning in favor of the LLM interpretation of the clinical text. This could have resulted in the synthetic images aligning well with the text interpretation and CCELLA class extraction but not necessarily with the radiologist re-classification of the corresponding real image, which could explain why \(S_R\) exhibited poorer performance. Thus, while the same train and test datasets were used for both LDM training and the downstream classifier task for consistency, future work should further explore the difference between real versus synthetic labeling for synthetic images when training a model for a medical task on an augmented dataset.

This study has a few limitations, each of which present an opportunity for future work. Firstly, this study emphasized data from a single institution, for a single MRI pulse sequence in a single pathology. While the pathology and institution restrictions were intentional to the study design in developing a limited-data LDM training pipeline, future work should seek to assess method generalizability by adapting this pipeline: a) to multi-institutional data, b) to images of different modality or pathology. Additionally, prostate MRI is clinically assessed in a minimum of three, and preferably four different MRI sequences~\cite{pesapaneComparisonSensitivitySpecificity2021,padhaniPIRADSSteeringCommittee2019}. Consequently, portions of the clinical text used for conditioning would have referred to MRI sequences that the LDM was neither trained on nor asked to generate. The use of one sequence alone in this study was chosen to focus primarily on the adapter innovation and data-limited pipeline design instead of the additional challenge of multi-sequence generation. Nonetheless, future work should seek to extend data-limited LDM training strategies for optimal multi-sequence generation. In particular, future work that incorporates diffusion weighted imaging would be important for clinical translatability. As well, this study included medical images from public datasets that did not have accompanying text reports for the purpose of gathering more training samples to train the denoising U-Net and the downstream classifier models. While outside the scope of this study, this prompts two opportunities for future work. One opportunity is to investigate how the ratio of conditioned to unconditioned data samples affect model training and whether there is a critical ratio below which training with conditioning becomes unstable or negatively affects training. Another opportunity is to explore synthetic clinical text report generation for these data samples. Finally, the intent of this study aligns with a growing body of literature seeking to use LDMs to generate synthetic medical data for the purpose of mitigating data scarcity challenges within machine learning for medical imaging applications. However, recent literature has also shown that LDMs can occasionally memorize and reproduce their training samples on inference~\cite{carliniExtractingTrainingData2023}. While outside the scope of this study, future work investigating and improving upon differential privacy within medical imaging LDMs would be valuable.

\section{Conclusion} 
This study presents CCELLA: a novel adapter for simultaneous alignment of nonmedical LLMs to the medical image space and radiology label extraction from medical text for LDM conditioning. This study also presents a CCELLA-centric pipeline for medical image LDM training given limited data volume and annotation. Using limited data and minimal human annotation, our proposed pipeline outperforms a recent medical LDM foundation model, and comparative baseline models from literature. Our proposed method also successfully augments the training dataset of a downstream classifier model for improved cancer prediction performance over training on real data alone in all classifier metrics, which was not consistently achieved by comparative medical LDMs. The results of this study suggest that CCELLA can successfully adapt pretrained, nonmedical LLMs for use in LDM conditioning for medical image generation, and can successfully train a 3D medical image LDM using data generated during clinical routine. Future work should focus on expanding this pipeline to multi-sequence image generation, validating the pipeline in a different medical imaging and pathology space, and exploring differential privacy within medical imaging LDMs.

\bibliographystyle{IEEEtran}
\bibliography{library}

\end{document}